\documentclass[final,5p,times,twocolumn]{elsarticle}
\usepackage{amsmath}
\usepackage{float}
\usepackage{tabularx}
\usepackage{tabu}
\usepackage{threeparttable}
\usepackage{dcolumn}
\usepackage{comment}
\usepackage{multirow}
\usepackage{braket}

\usepackage{graphics}
\usepackage{caption}
\usepackage{subcaption}
\usepackage{graphicx}

\usepackage{amssymb}
\usepackage{amsthm}
\usepackage[colorlinks=true]{hyperref}

\usepackage{natbib}
\makeatletter
\def\NAT@def@citea{\def\@citea{\NAT@separator}}
\makeatother

\biboptions{sort&compress,numbers}

\usepackage{color}

\journal{ ....}

\begin{document}

\begin{frontmatter}

\title{A Quantum Monte Carlo Study of mono(benzene)TM and bis(benzene)TM Systems}
\date{\today}
\author[a]{M.~Chandler~Bennett\corref{co1}}
\ead{mcbennet@ncsu.edu}
\author[a]{A.~H.~Kulahlioglu}
\cortext[co1]{Corresponding Author}
\author[a]{L.~Mitas}
\address[a]{Center for High Performance Simulation and Department of Physics, North Carolina State University, \\ Raleigh, North Carolina 27695-8202, USA}



\begin{abstract}
We present a study of mono(benzene)TM and bis(benzene)TM systems, where TM=\{Mo,W\}.
We calculate the binding energies by quantum Monte Carlo (QMC) approaches and compare the results with other methods and available experiments. 
The orbitals for the determinantal part of each trial wave function were generated from several types of DFT in order to optimize for fixed-node errors. 
We estimate and compare the size of the fixed-node errors for both the Mo and W systems with regard to the electron density and degree of localization in these systems. 
For the W systems we provide benchmarking results of the binding energies, given that experimental data is not available. 
\end{abstract}


\end{frontmatter}






\section{Introduction}
\label{sec:intro}

For the last several decades, quantum Monte Carlo (QMC) methods have been utilized to produce highly accurate electronic structure calculations of many-electron systems \cite{Kolorenc2011,bajdich2009,lester2009}. 
QMC methods offer an explicitly correlated wave function based alternative to more ubiquitous techniques that rely on expansions in determinants and basis sets to capture electron correlation.  
Diffusion Monte Carlo (DMC) has proven to be a particularly successful flavor of QMC whereby an initial wave function is evolved in imaginary-time to project out its ground state component, provided there exists a nonzero overlap between the two. 
Due to the fermion sign problem, however, a fixed-node approximation is generally made which constrains the nodal (hyper)surface (i.e., zero locus) of the evolving wave function to match that of an initial antisymmetric trial function. 
This restriction introduces a \textit{fixed-node bias} which only vanishes in the limit that the trial nodes match those of the exact fermionic ground state. 

\begin{figure}[hbtp]
   \centering
   \begin{subfigure}[c]{0.40\textwidth}
      \caption{\break}
      \includegraphics[width=\textwidth]{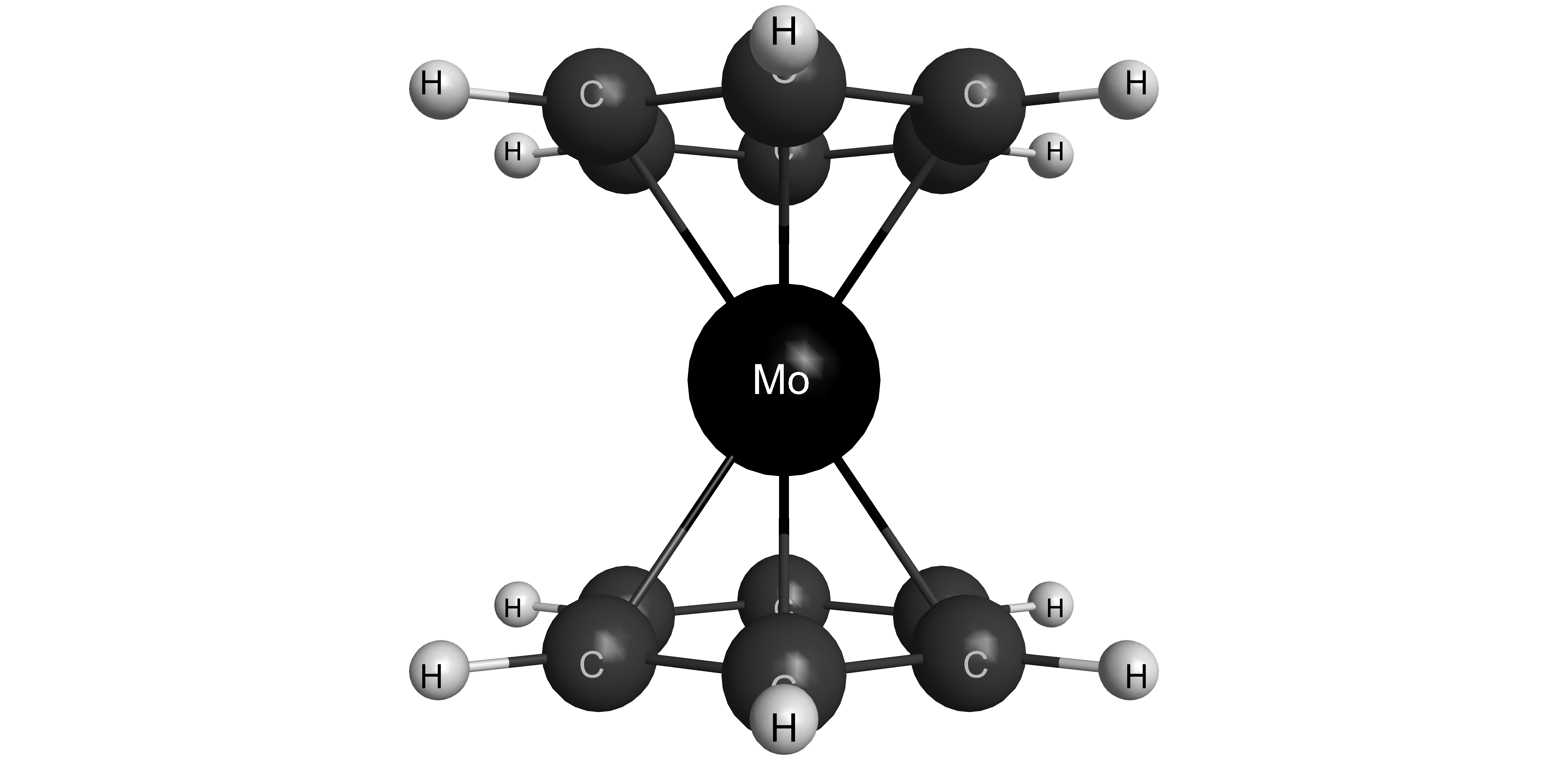}
      \break
      \label{fig:MoBz2}
   \end{subfigure}
   \centering
   \begin{subfigure}[c]{0.35\textwidth}
      \caption{}
      \includegraphics[width=\textwidth]{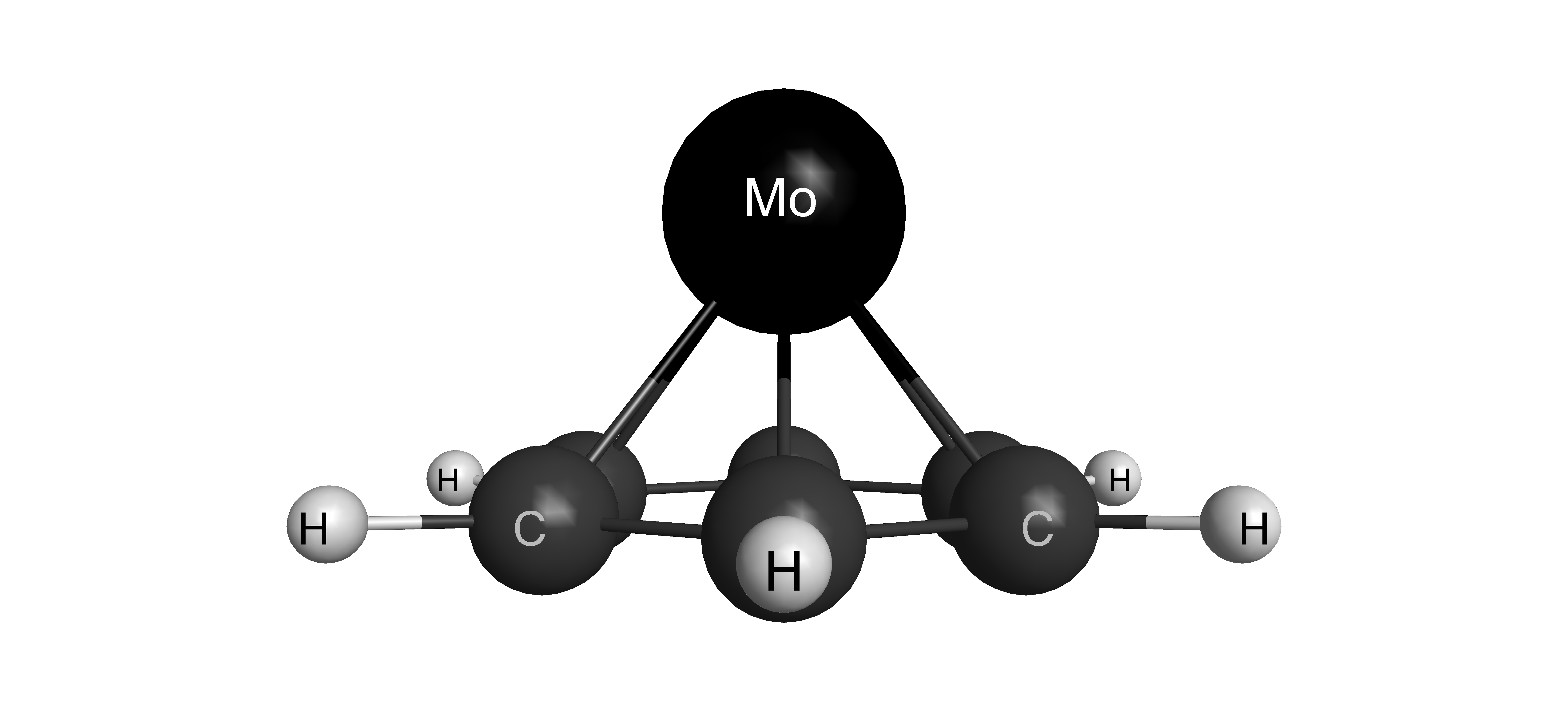}
      \label{fig:MoBz}
   \end{subfigure}
   \caption{Molecular geometries of (a) bis(benzene)molybdenum (MoBz$_2$) and (b) mono(benzene)molybdenum (MoBz). The geometries of the corresponding W systems are similar.}
   \label{bzmo01}
\end{figure}

The fixed-node bias is generally the dominant source of error in DMC calculations though accuracy in many cases can be quite high even for the simplest trial nodes, e.g., those from a single Slater determinant.
This nodal problem has been studied in various ways (see, for example, Refs. \cite{bajdich2005,bajdich2009}) and
several methods have been proposed to reduce the bias through some form of optimization of the trial nodes (see Refs. \cite{umrigar2007,toulouse2007a,luchow2007a}). 
Due to the stochastic nature of QMC, however, such optimizations are rather demanding on several fronts, in particular, the efficiency of the optimizer is vital and systematic improvements to the nodes can be difficult to obtain, especially for large systems.
To generate optimization methods that improve on these difficulties, a viable strategy would be to first obtain a better understanding of sources of nodal errors and of nodal properties, in general.
Some progress has been 
achieved in this respect in several studies \cite{Rasch2012,Rasch2014,Kulahlioglu2014}, where it was shown for $1^\text{st}$ and $2^\text{nd}$ row elements that nodal errors grow with node nonlinearities (a property related to the multiplicity of bonds) and with increasing electronic density. 
In this study, we add to this investigation by looking at nodal properties of benzene complexes containing molybdenum (Mo) and tungsten (W).
In particular, we consider the half-sandwich mono(benzene)Mo (MoBz) and the full-sandwich bis(benzene)Mo (MoBz$_2$) systems, depicted in Figure~\ref{bzmo01}, as well as the structurally and chemically similar WBz and WBz$_2$ systems.
The choice of these molecular systems has been motivated by several considerations.

Mo and W are $4d$ and $5d$ transition metal (TM) elements, respectively, and are stacked vertically within the periodic table's VIB column.
Given their partially filled $d$-shells, both elements have the potential to form many types of bonds and, therefore, can be found in variety of molecular systems that exhibit interesting properties with examples including catalysts and bioenzymes. 
As a consequence of these attributes, a comparison between Mo- and W-containing systems enables us to further examine the dependencies of fixed-node errors to electronic density (or, similarly, state localization) and bond multiplicity.
To achieve a sound comparison, we look at the same electronic states and 
similar geometries so that can we contrast the results between systems containing the $4^\text{th}$ period metal to those containing the $5^\text{th}$ period metal.

Another point of interest regarding these systems is in the fundamental chemistry of metals bonded to organic molecules.
As an example, CrBz$_2$ was first synthesized only a couple of decades ago and was the first stable synthetic structure of its kind, namely, where a TM was bound to two benzene rings in a full-sandwich geometry.
It was a breakthrough in the field of organometallic chemistry and was followed by many novel sandwich-type complexes such as VBz and CoBz. 
Some of these systems are well-known as catalysts since transition metals have the ability to form multiple bonds with low barriers and the structural flexibility of the sandwich geometries enable easy conformational changes that are crucial for efficient reaction paths. 
For example, the MoBz$_2$ system that we include in this study, is a well-known catalyst. 
Furthermore, the benzene-based organometallics later proved to be of high interest in the field of spintronics. 
In particular, benzene multi-decker nanowires with sandwiched transition metal atoms, such as V and Co, have become some of the most studied systems recently due to differences in conductivities in the two spin channels and their possible functioning as spin filters \cite{horvathova2012}.
Yet another area of research that involves these constituents is graphene doped with transition metal adatoms where benzene can serve as the simplest cluster model that can describe the local chemistry at the graphene hollow bonding site \cite{Nakada}.
The Mo and W systems we consider here provide interesting examples of organometallic bonding which is often quite challenging to describe by either the correlated or density functional theory (DFT) approaches. 

It is well-known that systems containing TM elements often pose significant challenges for experiments and simulations. 
These difficulties can be attributed to (i) strong many-body effects arising from the compact nature of the $d$-states resulting in significant
electronic correlation~\cite{Andersson1994,Dachsel1999,Hongo2012,Michalopoulos1982,Muller2009,Balasubramanian2002a,Balasubramanian2002,Borin2008,Efremov1978,Goodgame1982,Atha1982,Baykara1984,Zhang2004} and (ii) near degeneracy of the $d$-shell and outer $s,p$-shells which can generate a large number of low-lying hybridized states in molecular environments. 
In order to probe the electronic correlations and many-body effects we employ the FNDMC method that, apart from the fixed-node approximation, can provide very accurate results. 
In addition, relativistic effects impact the valence electronic structure and must be accounted for at some level if reasonable agreement with experiment is expected.
To account for relativity at some level, we have utilized scalar-relativistic effective core potentials, throughout.

\section{Methods and Computational Details} 
\label{sec:computational_details}

Fixed-node diffusion Monte Carlo (FNDMC) is a projector method that filters out the ground state, $\Phi_0$, of the symmetry class determined by
the nodal structure of a variationally optimized trial function, $\Psi_T$. 
This can be expressed as,
\begin{equation}
\Phi_0\propto\lim_{\tau \to \infty} e^{-\tau (H-E_T)}\Psi_T ,
\end{equation}
where $E_T$ is an offset tuned to the ground state energy, $E_0$, and $\tau$ (a real parameter) represents the progression through imaginary time. 
We can recast this equation into an importance sampled form given by the following convolution, 
\begin{equation}
f(\mathbf{R},\Delta\tau+\tau)=\int d\mathbf{R'} \, \tilde{G}(\mathbf{R'}\rightarrow\mathbf{R},\Delta\tau) \, f(\mathbf{R'},\tau)
\label{eq:conv}
\end{equation}
where  $f(\mathbf{R},\tau)=\Phi(\mathbf{R},\tau)\Psi_{T}(\mathbf{R})$ and the kernel is given by the corresponding Green's function.
$\mathbf{R}$ and $\mathbf{R'}$ denote the combined positions of all $N$ electrons in the system. 
One can show that the solution to Equation \ref{eq:conv} can be found by stochastic simulation. 
The sampling points are delta functions in 3$N$-dimensional space and they evolve according to the kernel/Green's function. 
Within a given time step $\Delta\tau$ the evolution involves diffusion, drift and proliferation/disappearance processes and in the long time limit one samples the stationary solution corresponding to the desired ground state. 
For antisymmetric states such a method runs into an exponential inefficiency due to the fermion sign problem. 
In order to avoid this problem we impose the fixed-node approximation which can be expressed as the following inequality
\begin{equation}
f(\mathbf{R},t)=\Phi(\mathbf{R},t)\Psi_{T}(\mathbf{R}) \geq 0 .
\end{equation}
The lowest energy solution with this constraint is obtained when the nodes of the solution $\Phi(\mathbf{R},t)$ are identical to the nodes of $\Phi(\mathbf{R},t)$. 
The fixed-node bias vanishes quadratically as the nodes of the trial wave function approach the exact nodes and consequently the quality of the trial wave function nodes is crucial in FNDMC. 
Although this is not immediately obvious the  nodal constraint enforced by the fixed-node approximation enables to employ the FNDMC method to calculate also
excited states. 

\subsection{Trial Wavefunctions}\label{theory4}
We employ trial wave functions of  the  Slater-Jastrow type  given by:
\begin{equation}
\psi_T(\mathbf{R})=\sum\limits_{i}c_i {\rm det}_i^{\uparrow}(\varphi_j(\mathbf{r}_k)){\rm det}_i^{\downarrow}
(\varphi_l(\mathbf{r}_m))e^{U(\mathbf{R})}
\end{equation}
where $c_i$ are expansion coefficients  and $U(\mathbf{R})$ is the Jastrow function. 
The nodes of a trial wave function are determined by the Slater determinants that are built from one-particle orbitals.  
The Jastrow function has one-body, two-body and three-body correlation terms given by
\begin{equation}
U(\mathbf{R})=\sum\limits_I\sum\limits_i U_1(r_{iI})+\sum\limits_{i > j}U_2(r_{ij})+ ...
\end{equation}
where i,j and I denote electron and nuclei, respectively, while $r_{iI}, r_{ij}$ are the corresponding distances. 
The functions $U_1, U_2, ... $ that capture the one-particle, two-particle, and so on, correlations are expanded in appropriate basis sets and optimized in variational Monte Carlo. 

After examining orbitals generated by different techniques such as Hartree-Fock (HF), natural orbitals from configuration interaction with single and double excitations (CISD) with varying size of virtual space and DFT orbitals, we found that the DFT orbitals provided lower energies, in general, and therefore were used in the FNDMC calculations. 

\subsection{Effective Core Potentials}\label{ecp}
We replace the atomic cores with effective core potentials (ECPs) since the QMC computational cost grows as $Z_{eff}^{5.5-6.5}$, where $Z_{eff}$ is the effective nuclear charge. 
ECPs therefore provide significant boost in efficiency. 
In addition, impact of relativity on valence states is built into the ECPs by construction so that relativistic effects at one-particle level are included. 
We note that the ECPs we use involve averaged spin-orbit interaction (often denoted as AREP, averaged relativistic effective potentials). 


ECP nonlocal operators were treated by the locality approximation \cite{Mitas1991}. 
The corresponding bias vanishes quadratically in the trial function error so that in many cases it gets folded into the fixed-node bias since it has the same type of scaling.  
The FNDMC calculations were done with  T-moves algorithm \cite{Casula2006} that preserve the variational bound for the total energy regardless of the localization approximation treatment of ECPs.    
The QMC calculations were performed using the software package \textsc{Qwalk}~\cite{Wagner2009}.

\section{Results and Discussion} 
\label{sec:results}

\subsection{MoBz and \texorpdfstring{MoBz$_2$}{MoBz\texttwosuperior} systems} 
\label{sub_sec:comp_details_MoBz_and_MoBz2}

Several functionals were examined for each system and the DFT calculations were carried out with the \textsc{Gaussian09}~\cite{g09} and \textsc{Gamess} \cite{Gamess1993} codes. 
The geometry optimizations were carried out within $C_{6v}$ and $D_{6h}$ for Mo-Bz and Mo-Bz$_2$, respectively. 
For Mo we used pseudopotentials with $4s4p4d5s$ valence space \cite{Peterson2007} while for the rest of atoms we used ECPs from Ref. \cite{Burkatzki2007}. 
The basis sets were of aug-cc-pVTZ quality.  

The spin multiplicity of the ground states of MoBz and MoBz$_2$ were examined in DFT, each yielding singlet ground states.
The trial wave functions used were the single-reference Slater-Jastrow as described above. 
The Slater determinant was constructed from single-particle orbitals generated by the DFT-TPSSh method that produced the lowest overall total energies, i.e., it led to the lowest fixed-node biases.
The Jastrow function in the trial wave functions contained up to three-body correlation terms, namely, nucleus-electron, electron-electron and nucleus-electron-electron terms. 
The Jastrow function was optimized within the framework of the VMC method.

\begin{figure}[t]
   \centering
   \begin{subfigure}[b]{0.5\textwidth}
      \caption{} 
      \includegraphics[width=\textwidth]{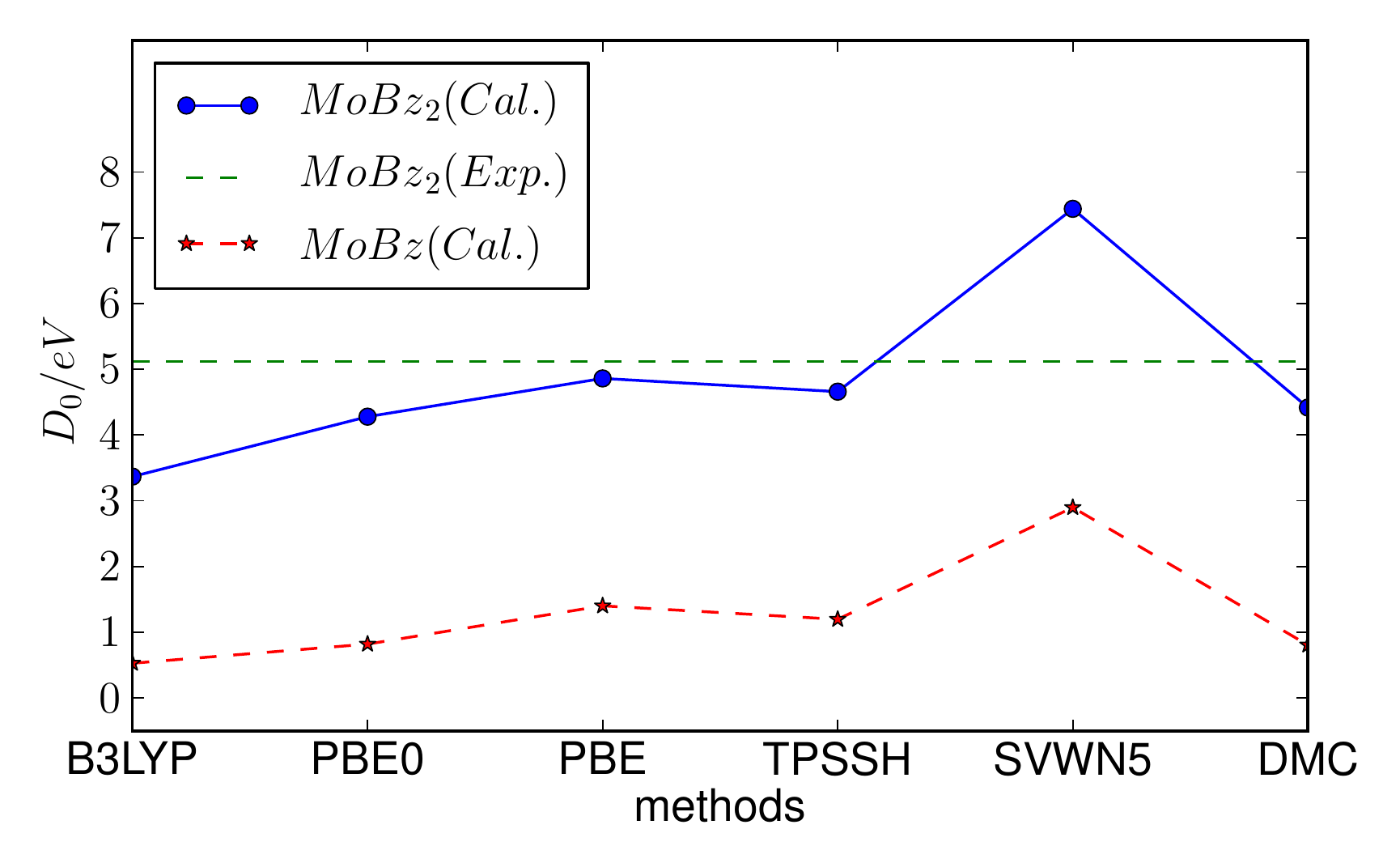}
      \label{fig:MoBz_binding}
   \end{subfigure}
   \vskip 3mm
   \centering
   \begin{subfigure}[b]{0.5\textwidth}
      \centering
      \caption{}
      \begin{tabular}{ccc}
         \hline \hline
         \textbf{Method} & \textbf{$D_0$/eV (MoBz$_2$)} & \textbf{$D_0$/eV (MoBz)} \\ 
         \hline
         B3LYP & 3.37 &0.53 \\
         PBE0 & 4.28 &0.82 \\
         PBE & 4.86 & 1.40 \\
         TPSSh & 4.66 & 1.20 \\
         SVWN5 & 7.43 & 2.90 \\
         DMC/TPSSh & 4.42(4) & 0.81(3) \\
         Exp. & 5.12~\cite{Sohnlein2006}&\\
         \hline \\
      \end{tabular}
      \label{tableb}
   \end{subfigure}
    \caption{The binding energies of the MoBz$_2$ and MoBz molecules calculated by DFT and FNDMC methods compared with experiment (a) and (b).
   FNDMC trial function has been constructed using TPSSh functional.}
   \label{bindingenergy_bzmo}
\end{figure}

The relaxed structural parameters used in the MoBz$_2$ and MoBz DMC calculations are given in Table~\ref{tab:Mo_Bz_bindings}.
The binding energies calculated by several DFT methods and the fixed-node DMC together with experiment are presented in Figure~\ref{bindingenergy_bzmo}. 
The binding energy is defined as follows:
\begin{align}
   D_0(\text{MoBz}_2)=&2E(\text{Bz})+E(\text{Mo})-E(\text{MoBz}_2) \\
   D_0(\text{MoBz})=&E(\text{Bz})+E(\text{Mo})-E(\text{MoBz}) 
\end{align}
Interestingly, for MoBz$_2$, all the DFT functionals except LDA/SVWN5 underbind with regard to the experimental value. 
An underbinding of about $0.66(3)$ eV is present also in the FNDMC result. 
We attribute most of this error to fixed-node error suggesting that the wave function has a significant multi-reference character. 
This is not surprising considering the high multiplicity of bonds in that system.  
A qualitatively similar pattern appears also for MoBz, however, in this case we were not able to find the experimental data. 
Our best guess is that the actual bonding energy lies around $1$ eV and that the FNDMC shows underbinding by $\approx 0.3$ eV, again due to the need of a multi-reference wave function to describe multiple bonds. 
The size of the fixed-node bias is similar to that in a recent study of corresponding dimers, in particular, the impact of the fixed-node errors on binding of the Mo dimer turns out to be about $\approx 0.7$ eV.

\begin{table}[t]
   \centering
   \caption{
   The structural parameters are given.
   The bond lengths ($R/$\AA) and dihedral angles ($\angle$/$^\circ$) were obtained from DFT-TPSSh calculations.
   }
   \begin{tabular}{ccccc}
      \hline \hline
      & $R$(Mo-Bz)  & $R$(C-C)  & $R$(C-H)  & $\angle$ CCCH \\ \hline
      MoBz$_2$  & 1.782     & 1.419   & 1.083  & 0.378        \\
      MoBz      & 1.617    & 1.438  & 1.093   & 3.463        \\
      \hline\\
   \end{tabular}
   \label{tab:Mo_Bz_bindings}
\end{table} 
%

\subsection{WBz and \texorpdfstring{WBz$_2$}{WBz\texttwosuperior} systems} 
\label{sub_sec:WBz_&_WBz2_Results}

For the benzene complexes containing W, we also carried out DFT calculations with various functionals using the \textsc{Gamess} \cite{Gamess1993} software package.
We again considered basis sets of aug-cc-pVTZ quality. 
An ECP was utilized for the W atom with a $5s5p5d6s$ valence space \cite{Peterson2007} and again ECPs from Ref. \cite{Burkatzki2007} were employed for the benzene constituents.
We imposed  $C_{6v}$ and $D_{6h}$ symmetries for WBz and WBz$_2$, respectively, and examined each system in a singlet spin state. 
The geometries of both systems were optimized with the quadratic approximation to the augmented Hessian technique until the maximum component of the gradient was less than $10^{-4}$ in magnitude. 
The equilibrated structural parameters are listed in tables \ref{tab:W-bz_geometry} and \ref{tab:W-bz2_geometry}, where we define the distance R(W-Bz) to be the distance from the W atom to the plane formed by the carbon atoms.

\begin{table}[!h]
    \centering
    \caption{Bond distances ($R/$\AA) and dihedral angles ($\angle/^\circ$) of WBz. }
    \begin{tabular}{lcccc}
        \hline\hline
              &   $R$(W-Bz) &    $R$(C-C) &    $R$(C-H) & $\angle$ CCCH    \\
        \hline
        PBE   & 1.589 & 1.442 & 1.092 & 3.717   \\
        PBE0  & 1.590 & 1.431 & 1.084 & 3.216   \\
        SVWN5 & 1.567 & 1.441 & 1.101 & 4.421   \\
        \hline
    \end{tabular}
    \label{tab:W-bz_geometry} 
\end{table}

\begin{table}[!h]
    \centering
    \caption{Bond distances ($R/$\AA) and dihedral angles ($\angle/^\circ$) of WBz$_2$.}
    \begin{tabular}{lcccc}
        \hline\hline
              &   $R$(W-Bz) &    $R$(C-C) &    $R$(C-H) & $\angle$ CCCH    \\
        \hline
        PBE   & 1.781 & 1.423 & 1.090 & 0.773   \\
        PBE0  & 1.773 & 1.413 & 1.082 & 0.760   \\
        SVWN5 & 1.748 & 1.422 & 1.099 & 1.218   \\
        \hline
    \end{tabular}
    \label{tab:W-bz2_geometry} 
\end{table}

For the FN-DMC calculations, single-reference Slater-Jastrow trial wave functions were used.
We considered Slater determinants built from PBE and PBE0 single-particle orbitals and up to three-body interaction terms in the Jastrow factors.  
The systems were placed in their respective optimized geometries.
The motivation for considering orbitals generated from the PBE functional in addition to the hybrid functional, PBE0, was to take the opportunity to investigate whether exact Hartree-Fock mixing in these systems would have any impact, as  was the case in $3d$ systems, see for example, \cite{Kolorenc2010}.
Table ~\ref{tab:W-bz_binding} shows the calculated binding energies for various DFT methods as well as fixed-node DMC; the latter we present as a benchmark result given the absence of experimental data for these systems.
The binding energies are defined similarly to those given in expressions (6) and (7), in addition to the following definition
\begin{equation}
   D_0(\text{WBz+Bz})=E(\text{WBz})+E(\text{Bz})-E(\text{WBz}_2).
\end{equation}
Provided that we only accounted for averaged spin-orbit interactions, we took the tungsten atom to be in its septet spin state which we determined to be the ground state of the Hamiltonian with spin-orbit averaged ECPs.  

\begin{table}[!h]
    \centering
    \caption{ Binding energies [eV] of W-Bz and W-Bz$_2$ and systems from DFT methods and from fixed-node DMC with two trial wave functions constructed with PBE and PBE0 orbitals.  
    } 
    \begin{tabular}{lccc}
        \hline\hline
         & W+Bz & W+Bz+Bz & WBz+Bz      \\
        \hline
        PBE        & 1.62        & 5.67       & 4.05  \\  
        PBE0       & 1.25        & 5.36       & 4.11  \\
        SVWN5      & 3.20        & 8.36       & 5.16  \\
        DMC/PBE  & 1.42(3)     & 6.00(3)    & 4.58(3) \\
        DMC/PBE0 & 1.47(3)     & 6.08(3)    & 4.61(3) \\
        \hline
    \end{tabular}
    \label{tab:W-bz_binding} 
\end{table}

\begin{figure}
\centering
\includegraphics[width=0.77\columnwidth,clip,angle=0]{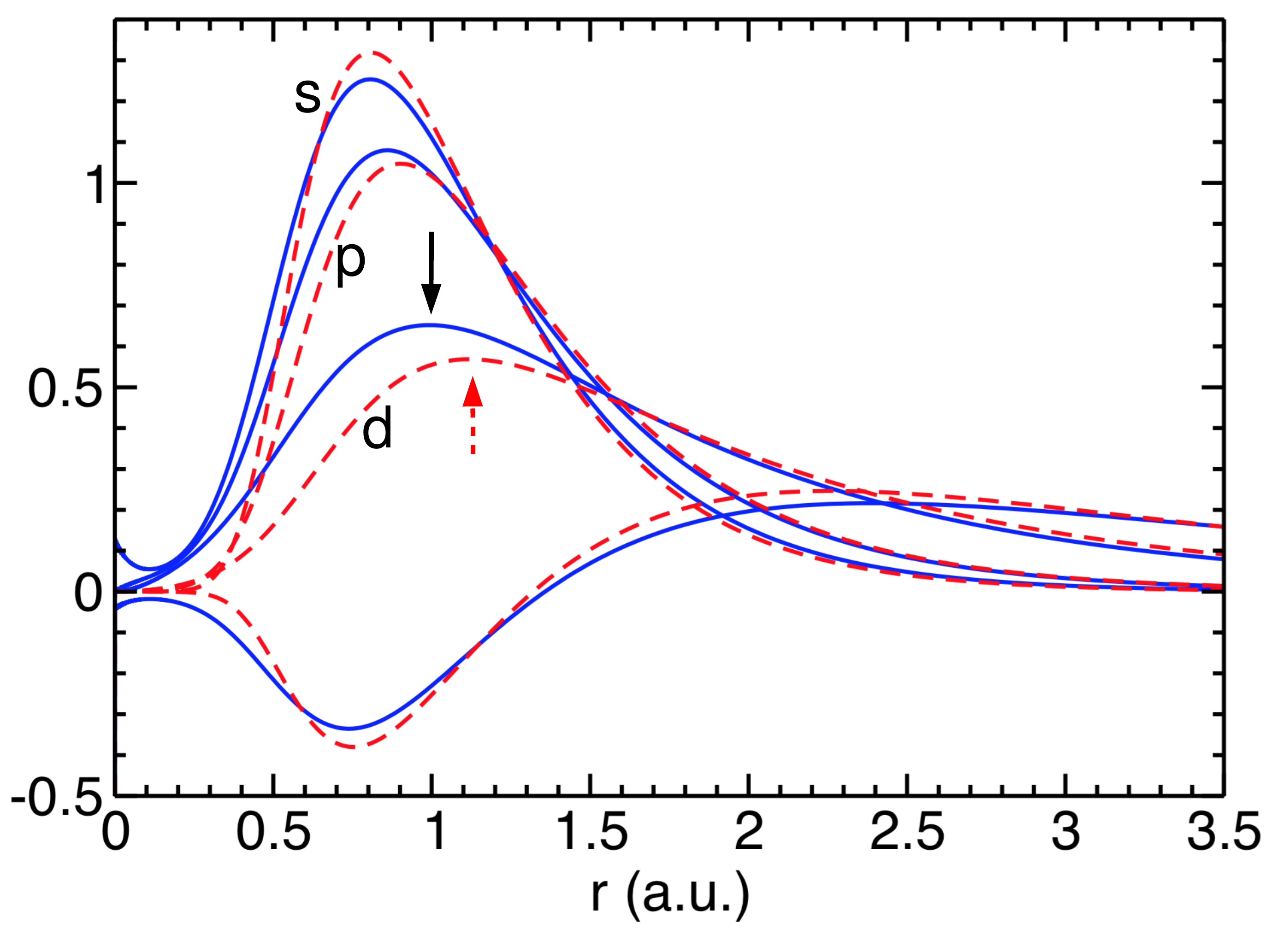}
\caption{Radial components of $s (\ell=0) $,  $p (\ell=1)$ and $d (\ell=2)$ valence pseudorbitals plotted as $r^{\ell} \varrho_{\ell}(r)$ for Mo (full, blue)  and W (dashed, red) atoms. The maximum of $s$-orbital for W is higher than for 
Mo (relativity), while the opposite is true in the $d-$channel. Also note the significantly
larger radius of the maximum (indicated by arrows) with the consequent smaller amount
of charge in the core region in the $d-$channel of W. }
\label{fig:csiorbs}
\end{figure}

It is interesting that the results for WBz and WBz$_2$ look rather different. 
Let us analyze the differences point by point.
First, we suspect that the part of the fixed-node bias that affects energy differences is significantly lower due to the lower electronic density and $5d$ states that are less localized when compared to 4$d$-states. 
This can be explained by inspection of the one particle atomic states plotted in Figure~\ref{fig:csiorbs}.
Note that the orbital maxima of the semicore $s$-states, namely, $4s$ for Mo and $5s$ for W, are almost the same. 
This contrasts with the $5d$ (Mo) and $6d$ (W) states that differ significantly. 
Not only is the maximum of the W $6d$ state at a larger radius but also the density inside the core region is much smaller. 
This is reminiscent of the contrast between the $2p$ states in the first-row and $3p$ states in the second-row where a 
much smaller fixed-node error was found in the latter case (both in absolute and in differences) \cite{Rasch2014}.

Second, the sensitivity of both FNDMC energy differences and total energies to changes in the DFT functionals used to generate the orbitals is very minimal
in the W systems, showing a reduced impact on the nodes. 
This conclusion is further supported by our study of W$_2$ that will be described elsewhere.  

Third, FNDMC binding energies are actually
larger than predicted by DFT functionals (except LDA, that significantly overbinds, as we commented above). 
Indeed, this is also in line with our calculations of W$_2$ that agrees with experiment rather well \cite{Bennett2016}. 
Therefore, we conjecture that our estimation of binding energies in these systems is the most accurate to date.
Note that we were not able to find reliable experimental data  for these systems in the literature and therefore we provide genuine predictions for these binding energies.
The fixed-node errors in Mo systems deserve further study as we well as the impact of the spin-orbit interactions that have been taken into consideration in an averaged manner in the present study. 
Overall, the FNDMC methods show very high accuracy for the $5\text{th}$ period systems in general and thus provide very accurate methodology for the correlated wave function calculations.

\section{Conclusions} 
\label{sec:conclusion}

We present calculations of single and double benzene systems with molybdenum and tungsten atoms in half- and full-sandwich geometries. 
We used several DFT functionals and FNDMC method to estimate the binding energies in these systems. 
For MoBz$_2$ we found that single-reference wave function appears to underestimate the binding by about 0.5 eV when compared with experiment. 
Interestingly, except LDA that is known to overbind by significant amount almost universally (20-30\%), the DFT functionals underbind these systems 
as well with significant spread 0.4-1.8 eV, depending on the functional. 
For MoBz we were not able to find the experimental value and therefore we estimate the binding to be around 1 eV. 
Since the overall behavior of the total ad binding energies is qualitatively similar to MoBz$_2$ we also estimate the fixed-node bias to cause 
underestimation of the FNDMC binding by about $\approx$ 0.4-0.6 eV with the uncertainty being assigned to the averaging of the spin-orbit effects. 
The results for tungsten systems appear to be significantly less impacted by the fixed-node errors due to the lower electronic density and more favorable spatial ordering of the $s,p$ and $d$ levels. This is a very encouraging finding since it implies that  accurate calculations are feasible for materials with $5d$ elements at the single reference trial function level. In the absence of experimental data, our calculations provide the most accurate values of binding energies in these tungsten molecular systems to date. 

{\bf Acknowledgments.} This research was supported by CDS\&E NSF Grant No. DMR-1410639. We also gratefully acknowledge XSEDE allocation at TACC.
We are very grateful for 
additional allocation on ANL Mira machine and for support from Dr. Anouar Benali. 





\bibliographystyle{science}

\bibliography{qmc_bz_systems}

\begin{thebibliography}{10}

\bibitem{Kolorenc2011}
J.~Koloren\v{c}, L.~Mitas, {\em Rep. Prog. Phys.}
  \href{http://dx.doi.org/10.1088/0034-4885/74/2/026502}{{\bf 74}, 026502}
  (2011).

\bibitem{bajdich2009}
M.~Bajdich, L.~Mitas, {\em Acta Physica Slovaca} {\bf 59}, 81-168 (2009).

\bibitem{lester2009}
W.~A. Lester, L.~Mitas, B.~L. Hammond, {\em Chem. Phys. Lett.} {\bf 478}, 1
  (2009).

\bibitem{bajdich2005}
M.~Bajdich, L.~Mitas, G.~{Drobn\'y}, L.~K. Wagner, {\em Phys. Rev. B} {\bf 72},
  075131 (2005).

\bibitem{umrigar2007}
C.~J. Umrigar, J.~Toulouse, C.~Filippi, S.~Sorella, R.~G. Hennig, {\em Phys.
  Rev. Lett.} {\bf 98}, 110201 (2007).

\bibitem{toulouse2007a}
J.~Toulouse, C.~J. Umrigar, {\em J. Chem. Phys.} {\bf 126}, 084102 (2007).

\bibitem{luchow2007a}
A.~{L\" uchow}, R.~Petz, T.~C. Scott, {\em J. Chem. Phys.} {\bf 126}, 144110
  (2007).

\bibitem{Rasch2012}
K.~M. Rasch, L.~Mitas, {\em Chem. Phys. Lett.}
  \href{http://dx.doi.org/10.1016/j.cplett.2012.01.016}{{\bf 528}, 59} (2012).

\bibitem{Rasch2014}
K.~M. Rasch, S.~Hu, L.~Mitas, {\em J. Chem. Phys.}
  \href{http://dx.doi.org/10.1063/1.4862496}{{\bf 140}, 041102} (2014).

\bibitem{Kulahlioglu2014}
A.~H. Kulahlioglu, K.~Rasch, S.~Hu, L.~Mitas, {\em Chem. Phys. Lett.}
  \href{http://dx.doi.org/10.1016/j.cplett.2013.11.033}{{\bf 591}, 170} (2014).

\bibitem{horvathova2012}
L.~Horv\'{a}thov\'{a}, M.~Dubeck\'{y}, L.~Mitas, I.~\v{S}tich, {\em Phys. Rev.
  Lett.} \href{http://link.aps.org/doi/10.1103/PhysRevLett.109.053001}{{\bf
  109}, 053001} (2012).

\bibitem{Nakada}
K.~Nakada, A.~Ishii, in J.~Gong (ed.), {\em Graphene Simulation}, pp. 3--15
  (2011).

\bibitem{Andersson1994}
K.~Andersson, B.~O. Roos, P.-{\r{A}}. Malmqvist, P.-O. Widmark, {\em Chem.
  Phys. Lett.} \href{http://dx.doi.org/10.1016/0009-2614(94)01183-4}{{\bf 230},
  391} (1994).

\bibitem{Dachsel1999}
H.~Dachsel, R.~J. Harrison, D.~A. Dixon, {\em J. Phys. Chem. A}
  \href{http://dx.doi.org/10.1021/jp982648s}{{\bf 103}, 152} (1999).

\bibitem{Hongo2012}
K.~Hongo, R.~Maezono, {\em Int. J. Quantum Chem.}
  \href{http://dx.doi.org/10.1002/qua.23113}{{\bf 112}, 1243} (2012).

\bibitem{Michalopoulos1982}
D.~L. Michalopoulos, M.~E. Geusic, S.~G. Hansen, D.~E. Powers, R.~E. Smalley,
  {\em J. Phys. Chem.} \href{http://dx.doi.org/10.1021/j100217a005}{{\bf 86},
  3914} (1982).

\bibitem{Muller2009}
T.~M\"{u}ller, {\em J. Phys. Chem. A}
  \href{http://dx.doi.org/10.1021/jp905254u}{{\bf 113}, 12729} (2009).

\bibitem{Balasubramanian2002a}
K.~Balasubramanian, {\em Chem. Phys. Lett.}
  \href{http://dx.doi.org/10.1016/S0009-2614(02)01500-2}{{\bf 365}, 413}
  (2002).

\bibitem{Balasubramanian2002}
K.~Balasubramanian, X.~Zhu, {\em J. Chem. Phys.}
  \href{http://dx.doi.org/10.1063/1.1497641}{{\bf 117}, 4861} (2002).

\bibitem{Borin2008}
A.~C. Borin, J.~P. Gobbo, B.~O. Roos, {\em Chem. Phys.}
  \href{http://dx.doi.org/10.1016/j.chemphys.2007.05.028}{{\bf 343}, 210}
  (2008).

\bibitem{Efremov1978}
Y.~M. Efremov, A.~N. Samoilova, V.~B. Kozhukhovsky, L.~V. Gurvich, {\em J. Mol.
  Spectrosc.} \href{http://dx.doi.org/10.1016/0022-2852(78)90109-1}{{\bf 73},
  430} (1978).

\bibitem{Goodgame1982}
M.~M. Goodgame, W.~A. Goddard, III, {\em Phys. Rev. Lett.}
  \href{http://dx.doi.org/10.1103/PhysRevLett.48.135}{{\bf 48}, 135} (1982).

\bibitem{Atha1982}
P.~M. Atha, I.~H. Hillier, {\em Molecular Physics}
  \href{http://dx.doi.org/10.1080/00268978200100231}{{\bf 45}, 285} (1982).

\bibitem{Baykara1984}
N.~A. Baykara, B.~N. McMaster, D.~R. Salahub, {\em Molecular Physics}
  \href{http://dx.doi.org/10.1080/00268978400101641}{{\bf 52}, 891} (1984).

\bibitem{Zhang2004}
W.~Zhang, X.~Ran, H.~Zhao, L.~Wang, {\em J. Chem. Phys.} {\bf 121}, 7717--7724
  (2004).

\bibitem{Mitas1991}
L.~Mitas, E.~L. Shirley, D.~M. Ceperley, {\em J. Chem. Phys.}
  \href{http://dx.doi.org/10.1063/1.460849}{{\bf 95}, 3467} (1991).

\bibitem{Casula2006}
M.~Casula, {\em Physical Review B}
  \href{http://dx.doi.org/10.1103/PhysRevB.74.161102}{{\bf 74}, 161102} (2006).

\bibitem{Wagner2009}
L.~K. Wagner, M.~Bajdich, L.~Mitas, {\em J. Comput. Phys.}
  \href{http://dx.doi.org/10.1016/j.jcp.2009.01.017}{{\bf 228}, 3390} (2009).

\bibitem{g09}
M.~J. Frisch, et~al., `Gaussian∼09 {R}evision {E}.01', Gaussian Inc.
  Wallingford CT 2009.

\bibitem{Gamess1993}
M.~W. Schmidt, et~al., {\em J. Comput. Chem.}
  \href{http://dx.doi.org/10.1002/jcc.540141112}{{\bf 44}, 1347} (1993).

\bibitem{Peterson2007}
K.~A. Peterson, D.~Figgen, M.~Dolg, H.~Stoll, {\em J. Chem. Phys.}
  \href{http://dx.doi.org/10.1063/1.2647019}{{\bf 126}, 124101} (2007).

\bibitem{Burkatzki2007}
M.~Burkatzki, C.~Filippi, M.~Dolg, {\em J. Chem. Phys.}
  \href{http://dx.doi.org/10.1063/1.2741534}{{\bf 126}, 234105} (2007).

\bibitem{Sohnlein2006}
B.~R. Sohnlein, D.-S. Yang, {\em J. Chem. Phys.}
  \href{http://www.ncbi.nlm.nih.gov/pubmed/16613453}{{\bf 124}, 134305} (2006).

\bibitem{Kolorenc2010}
J.~{Koloren\v c}, S.~Hu, L.~Mitas, {\em Phys. Rev. B} {\bf 82}, 115108 (2010).

\bibitem{Bennett2016}
M.~C. Bennett, C.~A. Melton, L.~Mitas, {\em to be published.} .

\end{thebibliography}







\end{document}